\documentclass[prb,showpacs,twocolumn]{revtex4-1}
\usepackage{graphicx}

\newcommand{\beq}{\begin{eqnarray}}
\newcommand{\eeq}{\end{eqnarray}}

\begin{document}

\title{Translation invariant tensor product states in a finite lattice system}

\author{J.W. Cai $^1$, Q.N. Chen$^2$, H.H. Zhao$^1$, Z.Y. Xie$^2$,  M.P. Qin$^1$, Z.C. Wei$^1$ and T. Xiang$^{1,2}$}

\address{
$^1$Institute of Physics, Chinese Academy of Sciences, P.O. Box 603, Beijing 100190, China}

\address{
$^2$Institute of Theoretical Physics, Chinese Academy of Sciences, P.O. Box 2735, Beijing 100190, China}

\begin{abstract}
We show that the matrix (or more generally tensor) product states in a finite translation invariant system can be accurately constructed from the same set of local matrices (or tensors) that are determined from an infinite lattice system in one or higher dimensions. This provides an efficient approach for studying translation invariant tensor product states in finite lattice systems. Two methods are introduced to determine these size-independent local tensors.
\end{abstract}

\maketitle

\section{Introduction}

Periodic boundary conditions are useful for simulating a large system by modeling a small part that is far away from its edges. In particular, a periodic system is free of boundary effects. It is easier to carry out finite size scaling analysis from the results obtained on periodic systems. This makes the extrapolation to the thermodynamic limit more transparent and smaller systems are needed in the simulations. Moreover, the energy-momentum dispersion of excitation states can be better studied in translation invariant periodic systems.

However, many numerical renormalization group methods, such as the density-matrix renormalization gropu (DMRG)\cite{White92}, works better in systems with open boundary conditions than those with periodic boundary conditions. This is because the DMRG wavefunction does not have the right entanglement structure in a periodic system. The criterion for the basis truncation in the DMRG is governed by the the bipartite entanglement entropy, which is bounded by the logarithm of the basis number $D$ retained in the truncation\cite{Eisert10}. To calculate the bipartite entanglement entropy, one needs to split the system into two parts by cutting one bond in an open chain, but two bonds in a periodic chain. Thus the entanglement entropy in a one-dimensional (1D) periodic system is about twice that in the corresponding open system. In the DMRG calculation with open boundary condition, the computational cost scales as $\mathrm{O}(LD^3)$ for a system of size $L$. However, to achieve the same accuracy, a periodic system needs roughly $D^2$ states per block. The computational cost scales roughly as $\mathrm{O}(LD^6)$. In two or higher dimensions, the entanglement entropy grows faster with the system size. The number of states needed scales exponentially with the number of boundary sites if periodic boundary condition is imposed.

The DMRG can be rephrased as a variational method over a class of matrix product states.\cite{Rommer95} The matrix product states and related algorithms have been actively explored. In 2004, Verstraete et al. pointed out that the matrix product wavefunction can also be used to cure the problem met in the DMRG with periodic boundary condition.\cite{Verstraete04} They proposed a variational approach to evaluate the matrix product state and showed that the local matrix elements can be determined by solving a generalized eigenvalue problem of dimension $D^2$. The computational cost of their algorithm  scales as $\mathrm{O}(LD^5)$.

Recently, Pippan et al. proposed an approximate scheme to evaluate the variational matrices used in the generalized eigenvalue equation in terms of a singular value decomposition\cite{Pippan10}. Their scheme reduces the computational effort from $\mathrm{O}(LD^5)$ to $\mathrm{O}(LD^3)$, which is comparable to that of DMRG with open boundary condition. More recently, Pirvu et al. showed that the efficiency for evaluating matrix product states with periodic boundary condition can be further improved if the system is translation invariant.\cite{Pirvu10} The cost of this method scales as $\mathrm{O}(mD^3)$, where $m$ is a number much smaller than $L$ and becomes a constant for very large system size. A $\mathrm{O}(D^3)$ cost can be also achieved by applying the variational Monte Carlo to a matrix product state.\cite{Sandvik07}

In this work, we further explore physical properties of translation invariant matrix or tensor product states in one or higher dimensions. We will show that one can use a set of size-independent local matrices to approximately but accurately represent the translation invariant ground states of all finite periodic systems. The performance of the algorithms for translational invariant systems explored in this work is improved over the other algorithms by dropping the size dependence in the calculation of the wave function. The ground state wavefunctions of all finite size systems can be obtained simply from the local matrices that are determined from an infinity lattice.

To understand this, let us consider the following translation invariant matrix product state in a 1D bipartite system
\begin{equation}\label{eq:mps1}
|\Psi \rangle  = \mathrm{Tr} (A[\sigma_1] B[\sigma_2] \cdots A[\sigma_{L-1} ] B[\sigma_L])|\sigma_1 \cdots \sigma_L\rangle ,
\end{equation}
where $|\sigma_i\rangle$ is the local basis state at site $i$ and $L$ is the lattice size. Given $\sigma$, $A[\sigma]$ and $B[\sigma]$ are $D\times D$ matrices. The trace in Eq.~(\ref{eq:mps1}) ensures periodic boundary condition. In the ground state, if one determines the local matrices by minimizing the energy, then $A$ and $B$ should in general be $L$ dependent. However, in many cases, such as in the valence bond solid state proposed by Affleck et al.\cite{Affleck88}, the local matrices can be $L$ independent. In the following, we will show that one can always use matrix product states with $L$-independent $A$ and $B$ to approximately but accurately represent the ground state wavefunctions in finite periodic systems.

The matrix product wavefunction of the ground state can be found by applying the projection operator  $\exp (-\beta H)$ to an arbitrary initial state $|\Psi_0 \rangle$.
\begin{equation}\label{eq:power1}
|\Psi_g \rangle = \lim_{\beta \rightarrow \infty} \exp ( -\beta  H )|\Psi_0 \rangle .
\end{equation}
This projection, which is equivalent to taking an imaginary time evolution, can be done iteratively. To do this, we divide $\beta$ into $N$ steps with a small incremental time $\tau = \beta /N$ and decompose the projection operator at each step using the Trotter-Suzuki formula
\begin{equation}
e^{-\tau H}= e^{-\tau H_{odd}} e^{-\tau H_{even}}  + o(\tau^2),
\end{equation}
where
\begin{eqnarray*}
H & = & H_{odd} + H_{even}, \\
H_{odd} & = & \sum_i h_{2i-1,2i} , \\
H_{even} & = & \sum_i h_{2i,2i+1} .
\end{eqnarray*}
At each iteration, the projection can be done in two successive steps, using the projection operators $\exp (-\tau H_{odd} )$ and $\exp (-\tau H_{even})$, respectively. Since all the terms in $H_{odd}$ or $H_{even}$ commute with each other, these projections
can be done by performing only \emph{local} operations. For example, by applying $\exp (-\tau H_{odd} )$ to Eq.~(\ref{eq:mps1}), the matrix product wavefunction will become
\begin{eqnarray}
|\Psi^\prime \rangle & = & e^{-\tau H_{odd}} |\Psi \rangle
\nonumber \\
& = & \mathrm{Tr}
\left( M[\sigma_1,\sigma_2] \cdots M[\sigma_{L-1}, \sigma_L] \right)|\sigma_1 \cdots \sigma_L\rangle ,
\end{eqnarray}
where $M[\sigma_1,\sigma_2] $ is a local matrix defined by
\begin{eqnarray}\label{eq:mps2}
&&  M_{\alpha \beta} [\sigma_1,\sigma_2]
 =  M_{\alpha\sigma_1 , \beta\sigma_2}
\nonumber \\
& = & \sum_{\sigma^\prime_1 \sigma^\prime_2 \gamma} A_{\alpha\gamma}[\sigma_1^\prime]B_{\gamma\beta}[\sigma^\prime_2] \langle \sigma_1,\sigma_2 | e^{-\tau h_{1,2}} | \sigma_1^\prime,\sigma_2^\prime \rangle .
\end{eqnarray}
By singular value decomposition, one can decompose $M$ to
\begin{equation}
M_{\alpha\sigma_1 ,\beta\sigma_2}= \sum_l U^M_{\alpha\sigma_1,l} \lambda^M_{l} V^M_{l,\beta\sigma_2} ,
\end{equation}
where $U_M$ and $V_M$ are unitary matrices, $\lambda_M$ is a semipositive vector. From this we can rewrite $M$ as a product of two matrices
\begin{eqnarray}
M[\sigma_1,\sigma_2] & = & A^\prime [\sigma_1 ] B^\prime [\sigma_2] ,
\label{eq:power2} \\
A^\prime_{\alpha, l} [\sigma_1 ] & = & U^M_{\alpha\sigma_1,l} \left( \lambda^M_{l}\right)^{1/2} ,
\\
B^\prime_{l,\beta} [\sigma_2] & = & \left( \lambda^M_{l}\right)^{1/2}
V^M_{l,\beta\sigma_2} .
\end{eqnarray}
$A^\prime [\sigma_1 ]$ is $D_1\times d D_2$ matrix and $ B^\prime [\sigma_2] $ is a $dD_2\times D_1$ matrix. Thus the wavefunction after the projection has the same matrix-product form as $|\Psi \rangle$. The only difference is that $A$ and $B$ in Eq.~(\ref{eq:mps1}) are now replaced by $A^\prime$ and $B^\prime$, respectively. Given $A$ and $B$, clearly the values of $A^\prime$ and $B^\prime$ are determined purely by the local Hamiltonian $h_{12}$ no matter how many sites in the system.

In the limit $\tau \rightarrow 0$ and $\beta = N\tau \rightarrow \infty$, the matrix product state obtained through above projection should approach the exact ground state. If there is no truncation to the matrix dimension, then the final site matrices $A$ and $B$ such obtained should be size independent. This means at least in the limit the bond dimension being infinity, the matrix product wavefunction of the ground state can be represented by the same local matrices, $A$ and $B$, no matter how large the system size is.

Of course, in practical calculation, one has to truncate the matrix dimension in order to carry out the projection for sufficiently many times. Otherwise, the matrix dimension will blow up exponentially with the projection steps. Upon truncation, the matrix product state  (\ref{eq:mps1}) with size-independent $A$ and $B$ will generally not be the rigorous ground state wavefunction. Nevertheless, it should still be a good approximation to the true ground state wavefunction. In particular, in the limit that the bond dimension approaches infinity, it should approach to the exact result. In the projection methods, the accuracy is limited only by the Trotter error controlled by $\tau$ and the truncation error controlled by the matrix dimension.

The above argument can be readily extended to the tensor-network states in two or higher dimensions. Tensor-network wavefunctions, for example, can be determined using the entanglement mean-field projection approach introduced in Refs.~[\onlinecite{Jiang08}] and [\onlinecite{Zhao10}]. Thus accurate ground state wave functions for all finite systems can be constructed by a few size independent local tensors. In higher dimensional systems, the lattice sizes that can be treated are generally very small due to the rapid growth of Hilbert space with the system size. The simplification of the problems to find a few tensors as suggested in this study is a big step towards solving the problems in two or higher dimensions.

Besides the projection method, the matrix product state in one dimension can be also obtained using several other methods\cite{Verstraete04,Pirvu10}. In an infinite lattice system, the local matrices obtained with different methods are equivalent. They can be gauged into a canonical form\cite{Garcia07} by certain unitary transformations.

\section{Determination of local matrices}

In this section, we introduce two approaches to evaluate the local matrices. One is an entanglement mean field projection approach. This is an approach that was first proposed for evaluating a matrix product state in an infinite lattice via the imaginary time evolution in 1D (Ref.~[\onlinecite{Vidal07}]) and 2D (Refs.~[\onlinecite{Jiang08,Zhao10}]). The other is the standard DMRG method with open boundary condition. This approach is  applicable only in 1D.

For the discussion below, we will take the Heisenberg spin chain as an example to show how accurate a translation invariant matrix product state with size independent local matrices can be. It is straightforward to extend the qualitative results to other quantum lattice models with short range interactions. The Hamiltonian of the Heisenberg model is defined by
\begin{eqnarray}
H =  \sum_i h_{i,i+1}, & \, & h_{i,i+1}  =  S_i \cdot S_{i+1} ,
\end{eqnarray}
where $S_i$ is the SU(2) spin operator.

\subsection{By projection}

Since the local matrices $A$ and $B$ are size independent, we can always use the methods that have been developed for studying a matrix product state in an infinite lattice to evaluate these matrices. In an infinite lattice, open boundary condition can be imposed without breaking the translation invariance of the matrix product state. In this case the two ends of the lattice are always disentangled and it is sufficient to perform just local transformation to convert a matrix product state into its canonical form. This can dramatically reduce the computational cost in the determination of local matrices.

In this regards, a commonly adopted approach is to take all matrix elements of $A$ and $B$ as variational parameters and determine them by minimizing the ground state energy. This, as discussed in Refs.~[\onlinecite{Verstraete04,Pippan10,Pirvu10}], can be achieved by solving a generalized eigenvalue problem. Since there is no entanglement between the two ends of the lattice, the dimension of the boundary matrix reduces to 1 and the cost of this algorithm\cite{Pippan10,Pirvu10} will scale as $D^3$.

Another kind of approach is to find the local matrices by performing the imaginary time evolution as defined by Eqs.~(\ref{eq:power1}-\ref{eq:power2}). These equations set a general framework for evaluating a matrix product state if no truncation is needed. However, in practical calculation, we have to truncate the matrix dimension at each step of projection. If we truncate the matrix dimension just using the singular value decomposition of $M$-matrix defined by Eq.~(\ref{eq:power2}), the matrix product state will generally not converge to the true ground state. This is because $M$ is purely a local matrix and the contribution from the environment matrix has not been considered in the basis truncation\cite{Zhao10,Xie09}.

The interplay between $M$ and the environment matrix can be properly and accurately handled by performing a number of transformations to enable $A$ and $B$ to satisfy the following canonical conditions
\begin{eqnarray}
\sum_\sigma A[\sigma] A^\dagger[\sigma] & = & I,
\\
\sum_\sigma B[\sigma] B^\dagger[\sigma] & = & I,
\\
\sum_\sigma A^\dagger[\sigma] \lambda_b^2 A[\sigma] & = & \lambda_a^2,
\label{eq:can1}\\
\sum_\sigma B^\dagger[\sigma] \lambda_a^2 B[\sigma] & = & \lambda_b^2,
\label{eq:can2}
\end{eqnarray}
where $\lambda_a$ and $\lambda_b$ are semipositive diagonal matrices. The square of the diagonal matrix element of $\lambda_a$ or $\lambda_b$ is the eigenvalue of the reduced density matrix for the two semi-infinity blocks separated by $AB$ or $BA$ bond. It measures the entanglement between these two blocks and is equal to the possibility of the corresponding basis vector in the matrix product state. The truncation can be done by just keeping the largest $D$ eigenvalues of $\lambda_{a,b}$ and the corresponding basis vectors. This can minimize the truncation error at each project step\cite{Orus08}, same as in the DMRG. The matrix product state such obtained is believed to be the most accurate one within the Trotter error. The cost of this algorithm also scales as $D^3$. But this method is not applicable to a tensor-network state in two or higher dimensions.

A more efficient and easy to implement approach, which will be used in the calculation below, is to evaluate the environment contribution by taking a mean field approximation. This is a generalization of the poor-man's approach of second renormalization of tensor-network states introduced in Ref.~[\onlinecite{Zhao10}]. The mean field parameters are the bond vectors that are introduced to approximately measure the entanglement between the matrix to be decomposed and rest of other matrices. No canonical transformation needs to be done in this algorithm. At each step of projection, the truncation error is larger than that obtained by the canonical transformation. However,  the truncation error is not accumulated in the process of iterations. The results obtained with this approach can reach the same accuracy as those obtained by the canonical transformation.\cite{Zhao10} This approach converges fast and is highly reliable, provided that the short imaginary time evolution operator is sufficiently close to a unitary operator.\cite{Vidal07} It can be used not just for studying matrix-product states in one dimension\cite{Vidal07}, but also for studying tensor-network states in two or higher dimensions\cite{Jiang08,Xie09,Zhao10}. A detailed introduction to this method can be found from Ref.~[\onlinecite{Vidal07}] for one-dimensional and Refs.~[\onlinecite{Jiang08,Zhao10}] for two-dimensional systems, respectively.

To consider the renormalization effect of environment on $M$-matrix, let us redefine $A$ and $B$ as
\begin{eqnarray}
A[\sigma ] & = & \Gamma^a[\sigma] \lambda_a, \\
B[\sigma ] & = & \Gamma^b [\sigma] \lambda_b,
\end{eqnarray}
where $\lambda_{a}$ and $\lambda_b$ are positive diagonal matrices (also called bond vectors) defined on the bonds connecting two neighboring sites. $\lambda_{a,b}$ do not depend on $\sigma$. In the canonical representation of the matrix product state, they are just the diagonal matrices defined in Eqs.~(\ref{eq:can1}) and (\ref{eq:can2}). In general, they can be considered as an approximation to the diagonal matrices in the canonical form. They measure the entanglement between the left and right blocks connected by $\lambda_a$ or $\lambda_b$ in an infinite system. The gauge degrees of freedom of the matrix product are also partially fixed by these bond vectors.

If the basis states of the left and right blocks connected to the $M$-matrix defined by Eq.~(\ref{eq:power2}) are orthonormal, as in a canonical form, then the environmental contribution to $M$ is proportional to $\lambda_b$ from left side of the environment and $1$ from the right side of the environment. Thus the entanglement field on the bond for $M$ is given by the singular values of the following environment modified $M$-matrix
\begin{eqnarray}
\tilde{M}[\sigma_1,\sigma_2]  & = &
\lambda_b M[\sigma_1,\sigma_2]
\nonumber \\
& = & \lambda_b \Gamma^a[\sigma_1] \lambda_a \Gamma^b[\sigma_2] \lambda_b .
\label{eq:emf1}
\end{eqnarray}
In the mean-field calculation, however, the basis states on both the left and right blocks are not orthogonalized. In this case, $\lambda_{a,b}$ is just an approximate measure of the entanglement. Nevertheless, it still provides an efficient and good account of the renormalization effect of the environment. The reason for this is that in the projection method, the truncation error is not accumulated and there is no need to minimize the truncation error at every step of iteration.

\begin{figure}[t]
\includegraphics[width=0.52\textwidth]{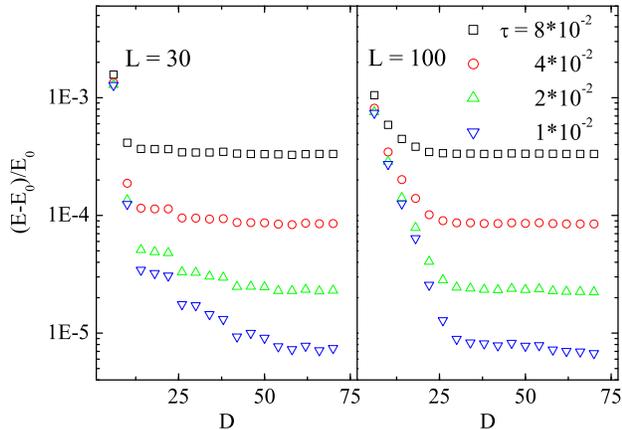}
\caption{The relative error of the ground state energy $(E-E_0)/E_0$ as a function of the matrix dimension $D$ for the spin half Heisenberg model with $L = 30$ and $L=100$ sites, respectively. $E_0$ is the exact ground state energy. The results for other system sizes are similar.}
\label{fig:half1}
\end{figure}

By SVD, we can decompose $\tilde{M}$ as a product of matrices
\begin{equation}
\tilde{M}_{\alpha,\beta}[\sigma_1,\sigma_2] \equiv \tilde{M}_{\alpha\sigma_1, \beta\sigma_2}
=  U_{\alpha\sigma, l} \Lambda_l V_{l,\beta \sigma_2} ,
\end{equation}
where $U$ and $V$ are unitary matrices and $\Lambda$ is a semipositive diagonal matrix. If $\tilde{M}[\sigma_1, \sigma_2]$ is $D\times D$ matrix for given $\sigma_1$ and $\sigma_2$, then $U$, $\Lambda$ and $V$ are all $dD\times dD$ matrices. To proceed the iteration, the matrix dimension needs to be truncated. This can be done by keeping the largest $D$ singular values of $\Lambda$ and the corresponding vectors. After this one can rewrite the matrix-product wavefunction back to the form defined by Eq.~(\ref{eq:mps1}) by updating $\Gamma^{a,b}$ and $\lambda_a$ with the formula
\begin{eqnarray}
\Gamma^a_{\alpha,l} [\sigma] & = & \lambda^{-1}_{b,\alpha\alpha} U_{\alpha\sigma, l} ,
\nonumber \\
\Gamma^b_{l,\alpha} [\sigma] & = & V_{l, \alpha,\sigma} \lambda^{-1}_{b,\alpha\alpha}  ,
\nonumber \\
\lambda_{a,ll} & = & \Lambda_{l,l} .
\nonumber
\end{eqnarray}
$\lambda_b$ remains unchanged. This completes one step of imaginary time evolution with $H_{odd}$. Similarly one can carry out the imaginary time evolution with $H_{even}$ and update all site matrices and bond vectors. The converged matrices from the iterated projections are used to construct ground state wavefunctions for all system sizes.

The error comes from two sources: one is the truncation error and the other is the Trotter error arisen from the Trotter-Suzuki decomposition. The Trotter error does not depend on the system size and can be reduced by using a smaller $\tau$ or using higher order Trotter-Suzuki decomposition formula. The truncation error is controlled by the matrix dimension $D$. It can be reduced by increasing the matrix dimension.

Fig.~\ref{fig:half1} shows the relative error of the ground state energy as a function of the matrix dimension $D$ for the spin-1/2 Heisenberg model on finite lattice systems obtained from the matrix product wavefunction with size-independent $A$ and $B$. We only show the results of relatively small system sizes, as the success of this method in the large system size limit is well documented. In the regime of small $D$, the truncation error dominates, the error drops quickly with increasing $D$. The error stops dropping when $D$ becomes larger than certain critical bond dimension $D_c$, beyond which the truncation error becomes smaller than the Trotter error. The value of $D_c$ depends on the small time interval $\tau$ and the order of Trotter-Suzuki decomposition formula used in the imaginary time evolution. But it does not depend much on the system size. This is probably because the same local matrices are used in the calculation of the ground state energy no matter how large the system size is. In obtaining the results in Fig.~\ref{fig:half1}, we have used the first order Trotter-Suzuki decomposition formula. In this case, the Trotter error scales approximately as $\tau^2$. This is consistent with the numerical results shown in Fig.~\ref{fig:half1}.

\begin{figure}[tb]
\includegraphics[width=0.52\textwidth]{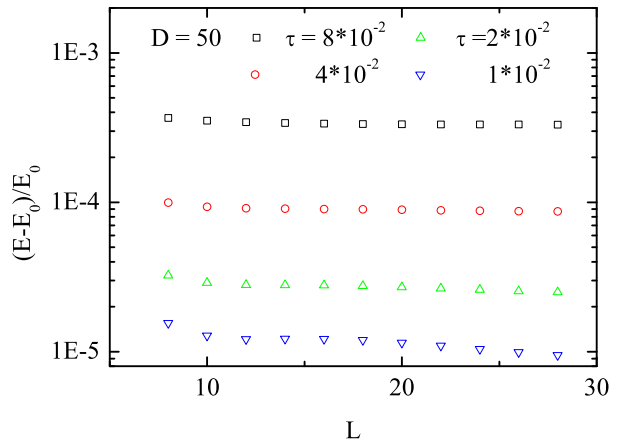}
\caption{The size dependence of the relative error of the ground state energy for the S=1/2 Heisenberg spin chain obtained with the matrix product state of bond dimension $D=50$.}
\label{fig:half2}
\end{figure}

Fig.~\ref{fig:half2} shows the lattice size dependence of the relative error of the ground state of the S=1/2 Heisenberg model with $D=50$. The values of $E_0$ are from the exact diagonalization results published in Ref.~[\onlinecite{Medeiros91}]. For the given $\tau$, $D=50$ are larger than the critical $D_c$, the errors are controlled by the Trotter error and  almost do not depend on the system size.

\begin{figure}[t]
\includegraphics[width=0.52\textwidth]{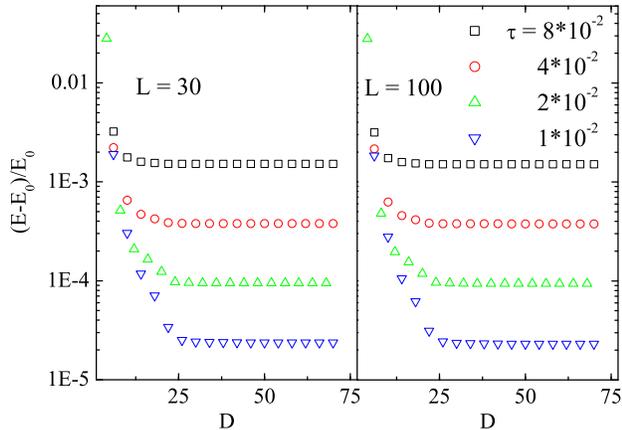}
\caption{The relative error of the ground state energy, $(E - E_0)/E_0$, as a function of matrix dimension $D$ for the spin one Heisenberg model. $E_0$ is the exact ground state energy.}
\label{fig:spin1}
\end{figure}

The matrix dimension dependence of the relative errors of the ground state energy behaves similarly for the S=1 Heisenberg spin chain, as shown in Fig.~\ref{fig:spin1}. Unlike the S=1/2 Heisenberg model whose excitation spectrum is critical, the S=1 Heisenberg model has a finite excitation gap and its ground state energy converges quickly with increasing $D$. Again the error is dominated by the truncation error in the small $D$ regime and by the Trotter error when $D$ is larger than a $\tau$-dependent critical $D_c$.

The above results show unambiguously that the matrix product states constructed with size-independent local matrices provide accurate and unified description of ground states irrespective of the system size. The accuracy of the wavefunction can be systematically and reliably improved by increasing the matrix dimension $D$ to reduce the truncation error and by using a smaller $\tau$ or higher order Trotter-Suzuki decomposition formula to reduce the Trotter error.

\subsection{By the DMRG}

In a translation invariant system, one can also use the DMRG method for open boundary systems to find the local matrices $A$ and $B$ defined by Eq.~(\ref{eq:mps1}). To understand this, let us consider the ground state generated by the DMRG with open boundary condition
\begin{equation} \label{eq:dmrg1}
| \Psi  \rangle  =  \sum_{\sigma_1 \sigma_2} \sum_{s_{0}e_{2}} \Psi (s_0, \sigma_1, \sigma_2,e_2 )|s_0 \sigma_1 \sigma_2 e_2 \rangle ,
\end{equation}
where we have used $1$ and $2$ to denote the coordinates of two middle sites of the chain, $|s_0\rangle$ and $|e_2\rangle$ are the basis states of the system and environment blocks, respectively. By singular value decomposition, we can decompose $\Psi (s_0, \sigma_1 ,\sigma_2 ,e_2 )$ as
\begin{eqnarray}
\Psi (s_0,\sigma_1,\sigma_2, e_2 ) & = &  \sum_l U^{(1)}_{ s_0 \sigma_1,l} \lambda^{(1)}_l  V^{(2)}_{l,\sigma_2 e_2}
\nonumber \\
& = &  \sum_l U^{(1)}_{ s_0 ,l}[\sigma_1] \lambda^{(1)}_l  V^{(2)}_{l,e_2} [\sigma_2] .
\label{eq:dmrg2}
\end{eqnarray}
In the above expression, the singular value $\lambda^{(1)}$ is nothing but the square root of the eigenvalue of the reduced density matrix. $U$ and $V$ are the unitary (or isometric after truncation) basis transformation matrices. The basis states of system and environment blocks are transformed according to the following equations
\begin{eqnarray}
|s_i \rangle & = & \sum_{s_{i-1} \sigma_i} U^{(i)}_{s_{i-1},s_i} [\sigma_{i}] |s_{i-1} \sigma_i \rangle,
\label{eq:dmrg3}
\\
|e_{i} \rangle & = & \sum_{e_{i+1} \sigma_{i+1}} V^{(i)}_{e_{i},e_{i+1}} [\sigma_{i+1}] | \sigma_{i+1} e_{i+1} \rangle .
\label{eq:dmrg4}
\end{eqnarray}

Substituting Eq.~(\ref{eq:dmrg2}) into Eq.~(\ref{eq:dmrg1}), we obtain
\begin{equation}
| \Psi  \rangle  =  \sum_{\sigma_1 \sigma_2} \sum_{s_0,e_2,l} U^{(1)}_{s_0,l}[\sigma_1] \lambda^{(1)}_{l} V^{(2)}_{l,e_2} [\sigma_2]
|s_0 \sigma_1 \sigma_2 e_2 \rangle .
\end{equation}
By using the basis transformation formula (\ref{eq:dmrg3}) and (\ref{eq:dmrg4}), we can further express this wavefunction as
\begin{eqnarray}
| \Psi  \rangle & = & \sum_{\sigma_0\sigma_1 \sigma_2\sigma_3} \sum_{s_{-1}s_0 e_2 e_3,l} U^{(0)}_{s_{-1},s_0}[\sigma_0] U^{(1)}_{s_0,l}[\sigma_1] \lambda^{(1)}_{l}
\nonumber \\
& &  V^{(2)}_{l,e_2} [\sigma_2] V^{(3)}_{e_2, e_3} [\sigma_3]
|s_{-1} \sigma_0 \sigma_1 \sigma_2 \sigma_3 e_3 \rangle .
\end{eqnarray}
If the system is reflection symmetric with respect to the middle bond of the chain, $V^{(2)}$ and $V^{(3)}$ should be the Hermitian conjugate of $U^{(1)}$ and $U^{(0)}$, respectively. In the DMRG iteration, the two local unitary matrices on the system block besides the middle bond, $U^{(0)}$ and $U^{(1)}$, will converge alternatively in a bipartite lattice model when the system size becomes sufficiently large\cite{Rommer97}. If we use these two converged isometric matrices to form a translation invariant matrix product state, i.e.
\begin{eqnarray}
A[\sigma] & = & U^{(0)} [\sigma ] ,
\label{eq:dmrg5}
\\
B[\sigma ] & = & U^{(1)} [\sigma ] ,
\label{eq:dmrg6}
\end{eqnarray}
we expect that it will be a good approximation to the ground state of the Hamiltonian with periodic boundary condition. This is because in an infinite system, the wavefunction in the middle of the chain should be the same no matter what kind of boundary condition is used.

\begin{figure}[t]
\includegraphics[width=0.52\textwidth]{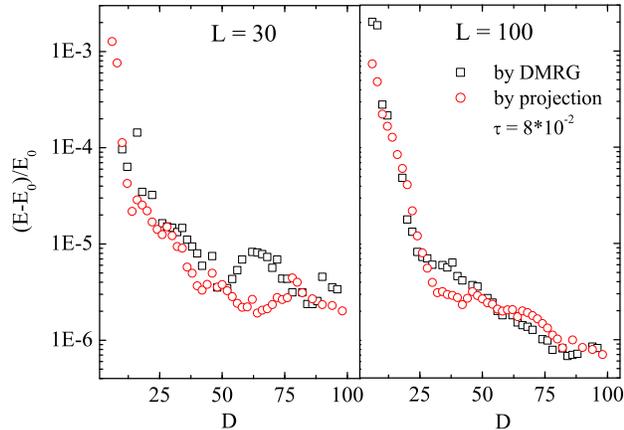}
\caption{Comparison of the relative error of the ground state energy of the spin-half Heisenberg model as a function of $D$ obtained using the matrix product states whose local matrices are determined by the DMRG with those determined by the poor-man's projection.}
\label{fig:dmrg}
\end{figure}

The local matrices such obtained are Trotter-Suzuki decomposition error free. Fig.~\ref{fig:dmrg} compares the relative error of the ground state energy for the $S=1/2$ Heisenberg model obtained by the DMRG method to that obtained by the entanglement mean-field projection method. In the entanglement mean-field projection calculation, we have chosen a small $\tau$ so that the Trotter error is much smaller than the truncation error. We find that the errors obtained with these two methods are of the same order.

To construct the translation invariant matrix product states using the DMRG is generally less efficient than the entanglement mean-field projection method. But the DMRG calculation can be combined with the conventional DMRG study with open boundary condition. No extra cost is needed in order to calculate the local matrices used by the matrix product state.

In the above discussion, we have assumed that the isometric matrices $U^{(0)}$ and $U^{(1)}$ would converge for sufficiently large chain. This is literally correct. However, if there are degeneracy or nearly degeneracy within computer machine error in the singular values $\lambda$ in Eq.~(\ref{eq:dmrg2}), $U$-matrices may not be uniquely fixed since the ground state is unchanged by swapping any pair of these degenerate states. Consequently the column index of $B[\sigma]$ may not perfectly match the row index of $A[\sigma]$ (there is no mismatch between the row index of $B[\sigma]$ and the column index of $A[\sigma]$). In this case, the local matrices defined by Eqs.~(\ref{eq:dmrg5}-\ref{eq:dmrg6}) may not be a good description of the ground state.

\section{Summary}

We have shown that the local matrices (tensors) obtained from an infinite lattice can be also used to accurately represent the matrix product states (or tensor-network states) in finite translation invariant lattice systems. This provides an efficient way to determine the matrix product states or tensor-network states in one- or higher-dimensional periodic systems.

The size-independent local matrices can be determined by the approximate entanglement projection with a poor-man's treatment to the environment lattice. This method can be applied not only in 1D, but also in two- or higher dimensions. The accuracy of the wavefunction such obtained is controlled by both the truncation and the Trotter-Suzuki decomposition errors. For small $D$, the truncation error dominates. This error, however, can be reduced below the Trotter-Suzuki decomposition error simply by increasing the bond dimension $D$. In that case, the accuracy of the matrix product wavefunction is purely determined by the Trotter-Suzuki decomposition error. This error can be reduced by using a small Trotter-Suzuki parameter $\tau$ or using a higher order Trotter-Suzuki decomposition formula. One can also find the size-independent local matrices using the DMRG with open boundary conditions. The matrix product states such obtained do not have the Trotter-Suzuki decomposition error.

We acknowledge the support of NSF-China and the National Program for Basic Research of the Ministry of Science and Technology of China.


\begin{thebibliography}{99}

\bibitem{White92} S. R. White, Phys. Rev. Lett. {\bf 69}, 2863 (1992).

\bibitem{Eisert10} J. Eisert, M. Cramer and M. B. Plenio, Rev. Mod. Phys. {\bf 82}, 277 (2010).

\bibitem{Rommer95} S. Ostlund and S. Rommer, Phys. Rev. Lett. {\bf 75}, 3537 (1995).

\bibitem{Verstraete04} F. Verstraete, D. Porras, and J. I. Cirac, Phys. Rev. Lett. {\bf 93}, 227205 (2004).

\bibitem{Pippan10} P. Pippan, S. R. White, and H. G. Evertz, Phys. Rev. B {\bf 81}, 081103(R) (2010).

\bibitem{Pirvu10} B. Pirvu, F. Verstraete and G. Vidal, arxiv:1005.5195v1 (2010).

\bibitem{Sandvik07} A. W. Sandvik and G. Vidal, Phys. Rev. Lett. {bf 99}, 220602 (2007)


\bibitem{Affleck88} I. Affleck, T. Kennedy, E. H. Lieb, and H. Tasaki, Commun. Math. Phys. {bf 115}, 477 (1988).


\bibitem{Jiang08} H.C. Jiang, Z.Y. Weng, T. Xiang, Phys. Rev. Lett. {\bf 101}, 090603 (2008).

\bibitem{Zhao10} H. H. Zhao, Z. Y. Xie, Q. N. Chen, Z.C. Wei, J.W. Cai, and T. Xiang, Phys. Rev. B {\bf 81}, 174411 (2010).

\bibitem{Garcia07} D. Perez-Garcia, F. Verstraete, M.M. Wolf, J.I. Cirac, Quantum Inf. Comput. {\bf 7}, 401 (2007).


\bibitem{Vidal07} G. Vidal, Phys. Rev. Lett. {\bf 98}, 070201 (2007).

\bibitem{Xie09} Z.Y. Xie, H. C. Jiang, Q. N. Chen, Z.Y. Weng, and T. Xiang, Phys. Rev. Lett. {\bf 103}, 160601 (2009).

\bibitem{Orus08} R. Orus and G. Vidal, Phys. Rev. B {\bf 78}, 155117 (2008).

\bibitem{Medeiros91} D. Mederiros and G. G. Cabrera, Phys. Rev. B {\bf 43}, 3703 (1991).

\bibitem{Rommer97} S. Rommer and S. Ostlund, Phys. Rev. B {\bf 55}, 2164 (1997).

\end{thebibliography}
\end{document}